\documentclass[12pt,preprint]{aastex}
\pdfoutput=1

\slugcomment{Astro-ph preprint of AJ article}

\shorttitle{Interpretation of Stephan Quintet}
\shortauthors{Gibson \& Schild}

\begin{document}


\title{Interpretation of the Stephan Quintet Galaxy Cluster using
Hydro-Gravitational-Dynamics: Viscosity and Fragmentation}


\author{Carl H. Gibson\altaffilmark{1}}
\affil{Departments of Mechanical and Aerospace Engineering and 
Scripps Institution
of Oceanography, University of
California,
     San Diego, CA 92093-0411}

\email{cgibson@ucsd.edu}

\and

\author{Rudolph E. Schild}
\affil{Center for Astrophysics,
     60 Garden Street, Cambridge, MA 02138}
\email{rschild@cfa.harvard.edu}


\altaffiltext{1}{Center for Astrophysics and Space Sciences, UCSD}


\begin{abstract}
Stephan's Quintet (SQ) is a compact group of galaxies that has been well
studied since its discovery in 1877 but is mysterious using cold
dark matter hierarchical clustering cosmology  (CDMHCC). Anomalous red shifts $z
= (0.0027,0.019, 0.022,  0.022,  0.022)$ among galaxies in SQ either reduce it
to a Trio with two highly improbable intruders from CDMHCC or support the
Arp (1973) hypothesis that its red shifts are intrinsic.  An alternative is
provided by the Gibson 1996-2006 hydro-gravitational-dynamics  (HGD) theory where
superclusters, clusters and galaxies all originate by gravitational fragmentation in the super-viscous plasma epoch and at planetary and star cluster mass scales in the primordial gas of the expanding universe. 
By this fluid-mechanical cosmology, the SQ galaxies
gently separate and  remain precisely along a line of sight because of
perspective and the small transverse velocities permitted by their sticky
viscous-gravitational beginnings.  Star and gas bridges and 
young-globular-star-cluster (YGC) trails  observed by the Hubble Space Telescope are triggered as
SQ galaxies separate through viscous baryonic-dark-matter halos  of
dark proto-globular-cluster (PGC) clumps of frozen Earth-mass
primordial-fog-particles (PFPs). 
\end{abstract}




\section{Introduction}

Stephan's Quintet (SQ, HCG 92, Arp 319, VV 288) is one of the first known
(Stephan 1877) and best studied of the Hickson 1982 catalog of very compact
groups of galaxies, and increasingly the most mysterious.  The group consists
of the Trio NGC 7319, NGC 7318A, and NGC 7317, 
all of which have precisely the same redshift
0.022, NGC 7318B with redshift 0.019 closely aligned with the Trio member NGC 7318A, and NGC
7320 with  $z = 0.0027$.  
Burbidge and Burbidge 1959 noted that the large discrepancy of redshifts
for the double galaxy NGC 7318AB would require very large mass/light ($M/L$) ratios
$\approx 300 \pm 200$ from dynamical models to achieve virial equilibrium. 
However, the true mystery of SQ began when the missing redshift for NGC 7320
 was determined by Burbidge and Burbidge 1961 to be a mere $z = 0.0027$, with
relative velocity $cz = 8.1 \times 10^{5}$ m s$^{-1}$ compared to $  6.7 \times
10^{6}$ for the Trio.  For virial equilibrium, this increases the kinetic
energy of the group by a factor 
$\sim 30$ and would require $M/L \approx 10,000$: much too large to be
credible.  Thus it was concluded 
\citep{bur61} that the system must be in a state of explosive expansion since
the $\it a$ $ \it priori$ chance of NGC 7320 not being a member of the group but
a random foreground galaxy  is about 1/1500.  A connecting gas bridge to the Trio  \citep{gut02} confirms that NGC 7320 is not a chance intruder but a separated companion.  This
conflicts with cold-dark-matter hierarchical-clustering (CDMHC) cosmology where galaxy clusters
form by condensation, not fragmentation.  

An alternative cosmological proposal is that the SQ galaxy redshifts are 
simply variable by unknown physics after ejection from the same parent-galaxy
active-galactive-nucleus (AGN) and remain at the same distance closely clumped.   Arp 1973
summarizes several of his papers of 1970-1972 from which he concludes that the AGN of the 
nearby large spiral galaxy NGC 7331 ($z = 0.0027$) has ejected all of the SQ galaxies, some with
intrinsic red shifts, so that the SQ galaxies are all located at the same
$\approx 9$ Mpc ($3\times 10^{23}$ m) distance of their parent NGC 7331 and NGC 7320, not at 74 Mpc and 64 Mpc implied by red shifts of the Trio and NGC 7318B, respectively.  Arp lists numerous
cases where galaxies in close angular proximity not only have widely different
red shifts but have coincident spin magnitudes and alignments with observed AGN jets,
consistent with his claim that the ejection of galaxies and quasars from AGNs with intrinsic red shifts is  amply justified by the accumulation of circumstantial evidence and the lack of an alternative hypothesis that fits all the data
\citep{arp98}.  

In support of the Arp proposal, galaxies and quasars 
frequently show evidence of ejection with intrinsic red
shifts
\citep{hoy00}.  A Seyfert 1 galaxy (NGC 6212) is observed closely surrounded by
a large number  ($\ge$44) of QSOs that it may have ejected \citep{bur03}, with
QSO surface densities (69 per square degree) larger than ambient by estimated
factors of 30 to 10 and  decreasing  (to 17) with angular
distance for radii 10$\--$50 minutes.   It has  been suggested
\citep{hoy00} that the big bang hypothesis itself may be questioned based on
the remarkable accumulation of such coincidences that are contrary to the
statistics of standard CDM hierarchical galaxy clustering cosmology (CDMHCC)
and the Hubble red-shift  radial-velocity relationship ($v = cz$) of big bang
cosmology.  

These mysteries vanish when the observations are interpreted using the
hydro-gravitational-dynamics theory (HGD) of Gibson 1996-2006.  From HGD cosmology
the HCG92-SQ anomalies
 are simply fossil manifestations of the viscous-gravitational beginnings of galaxy clusters and galaxies by fragmentation early in the plasma epoch \citep{gib96}.  The apparent close proximity of the SQ galaxies is an optical illusion resulting from perspective and the small transverse 
 velocities permitted by strong  friction from the increasingly sticky
  non-barionic dark matter (freezing earth-mass H-He planets in 
  proto-globular-star-cluster clumps) as it slowly
diffuses from a late-fragmenting dense linear galaxy-cluster reflecting Nomura scale proto-galaxies
 \citep{gib00, gib06}. Thus dense angular
galaxy clusters may have wide spatial separations, where their galaxies lie precisely along a line of
sight from their sticky origins and sticky dark matter and the uniform expansion of space expected from big bang turbulence cosmology \citep{gib05,gib04}.  A nearby member of a fragmenting compact
cluster affected only by universe expansion will show its origin directly behind it.
From the large range of redshifts (0.03 to 2.6) and Hubble distances (150 to
3730 Mpc) of the NGC 6212-quasar system
\citep{bur03} the observed
AGN-QSO galaxies are concentrated in a
thin  line-of-sight pencil ($\approx$150/1), contradicting CDMHCC and
supporting HGD.  The pencil containing the Stephan Quintet galaxies is even thinner, with L/D $\approx$1000/1.   

 CDMHC cosmology  is increasingly in conflict with hydrodynamic theory and with observations \citep{gib05, gib04, gs07}.   It should be abandoned.   Galaxies from HGD cosmology begin gently at the plasma-gas transition with dense clusters of proto-galaxies already formed by fragmentation with
 sizes and geometry reflecting the strain-rates and viscosity of the weak plasma turbulence.  The
 large decrease in viscosity on transition dramatically decreased fragmentation scales and triggered
 the first condensations leading to frozen dark matter planets in PGC clumps.
Stars formed immediately (with no dark ages) from binary 
mergers of the original hot planet-mass clouds.

  CDMHC cosmology begins violently much later with huge unstable Population III stars that re-ionized the gas with enormous light intensities that can now be ruled out \citep{ah06}.  CDM condensations based on the Jeans 1902 theory are unstable \citep{gib06} and will disintegrate into particles from tidal forces rather than clustering as assumed by CDMHC cosmology.  The ``dark energy'' hypothesis of $\Lambda \rm CDMHC$ is an artifact
 of supernova Ia events dimmed by evaporated dark matter planet atmospheres \citep{gs07}.  Dark
 energy and $\Lambda$ should be abandoned along with CDMHC.

In the following, HGD theory is reviewed  ($\S2$) and used to interpret and compare observations of SQ from the Hubble Space Telescope  ($\S3$) at optical frequencies \citep{gal01} with ground based telescope observations  ($\S4$)
 in the R and H$\alpha$ bands \citep{gut02}.  From the observations, clear evidence is found showing the SQ cluster red shifts are due to gentle separations of densely clustered galaxies from the expansion of the universe, contradicting CDMHC cosmology and the Arp hypothesis of intrinsic red shift galaxies ejected by AGNs.  From HGD, interpretations of compact groups [CG] and fossil groups [FG] of galaxies as mergers
 \citep{men07} should be reconsidered.  Conclusions are summarized ($\S5$).

\section{Hydro-Gravitational-Dynamics Theory}

Standard CDMHC cosmologies are based on highly over-simplified fluid
mechanical equations that assume the fluid is
collisionless and linear and that confuse hydrodynamics with hydrostatics
to claim incorrectly that gravitational structure formation on scales smaller than the Jeans scale
will be prevented by ``pressure support'' \citep{gs07}.  
The Jeans 1902 theory neglects viscous forces, 
turbulence forces, non-acoustic density fluctuations and important effects of
diffusion on gravitational structure formation.  Jeans did linear perturbation
stability analysis (neglecting turbulence) of Euler's equations (neglecting
viscous forces) for a
nearly uniform ideal gas with density
$\rho$ only a function of pressure (the barotropic assumption),  which reduced
the problem of gravitational instability to the solvable equations of
gravitational acoustics.  To reconcile his equations with the linearized
collisionless Boltzmann's equations and the resulting Poisson's equation for the
gravitational potential, Jeans assumed the density $\rho$ was zero.  This
assumption is known as the ``Jeans swindle''.    The only critical
wave length for gravitational stability with all these 
questionable assumptions (``swindles'')
is the Jeans length scale
$L_J$ where
\begin{equation}  L_J \equiv V_S/(\rho
G)^{1/2}
\approx (p/\rho^2 G)^{1/2} ,
\label{eq1}
\end{equation}
  $G$ is Newton's
gravitational constant and
$V_S \approx (p/\rho)^{1/2}$ is the sound speed.

Density fluctuations in fluids are not barotropic 
as assumed by
Jeans 1902 except rarely in small regions for short times near powerful sound
sources.  Density fluctuations that triggered the first gravitational
structures in the primordial fluids of interest were likely
non-acoustic (non-barotropic) density variations from turbulent mixing of
temperature or chemical species  concentrations produced by the big bang
 \citep{gib01,gib04, gib05,gib06}
 as shown by turbulent signatures in the cosmic microwave
background temperature anisotropies \citep{bs02}.  

Without viscous and turbulent forces or diffusion, fluids with non-acoustic
density fluctuations are absolutely unstable to the formation of structure due
to self gravity
\citep{gib96}.  Turbulence or viscous forces can dominate gravitational forces
at small distances from a point of maximum or minimum density to prevent
gravitational structure formation, but gravitational forces will dominate
turbulent or viscous forces at larger distances to cause structures if the gas
or plasma does not diffuse away faster than it can condense or rarify due to
gravity \citep{gib00}.  

The concepts of pressure support and thermal support reflect
a failure to distinguish fluid dynamics from hydrostatics.  Pressure forces
cannot prevent gravitational structure formation in the plasma epoch because
pressures equilibrate according to Bernoulli's equation
 in time periods smaller that the gravitational free fall
time $(\rho G)^{-1/2}$ on length scales smaller than the Jeans scale.  The
 Jeans scale in the primordial plasma is larger than the Hubble
scale of causal connection $L_H = ct$, where $c$ is light speed and $t$ is
time.  Information about density variations needs time to be transmitted.  The initial
stages of collapse on a density maximum or fragmentation at a density minimum are
quite gentle, so isentropic and adiabatic assumptions are justified \citep{gib00}.   Because
the early universe rapidly expanded with rate-of-strain $\gamma \approx t^{-1}$, gravitational 
structure formations inhibited by viscosity and turbulence
 were  exclusively by fragmentation until after the end of the plasma epoch \citep{gib96}.

Bernoulli's equation 
expresses the first law of thermodynamics during gravitational structure formation.  
Because energy losses by viscous friction are negligible in the initial stages, 
the enthalpy $p/\rho$ decreases
 to exactly match increases in kinetic energy $v^2 /2$ during gravitational condensation or fragmentation so the Bernoulli group $B = p/\rho + v^2 /2 $ remains constant and the term
 grad$B$ in the momentum equation is zero.  As the speed increases the pressure goes down, which is the wrong sign for any pressure support.  Because pressure only appears in the Bernoulli group
 in the Navier Stokes momentum equations, there is no pressure support term.  Pressure support only occurs in hydrostatics; that is, when the fluid velocity is near zero and pressure gradient forces balance
 gravitational forces.

Therefore, if hydrodynamic gravitational forces exceed viscous and turbulence forces
in the plasma epoch at scales smaller than $L_H$ then gravitational structures
will develop, independent of the Jeans criterion.  Only a very large
diffusivity of the plasma ($D \gg \nu$) could interfere \citep{gib00}.  The
diffusion velocity is
$D/L$ for diffusivity
$D$ at distance $L$ and the gravitational velocity is $L \rho^{1/2} G^{1/2}$.
The two velocities are equal at the diffusive Schwarz length scale
\begin{equation} L_{SD} \equiv
[D^2 / \rho G]^{1/4}.\end{equation}  

Thus very weakly collisional particles such
as the hypothetical cold-dark-matter (CDM) material cannot form 
potential wells
for baryonic matter collection because the particles have large diffusivity
and will disperse, consistent with observations \citep{sa02}.  Diffusivity
$D 
\approx V_p \times
L_c$, where $V_p$ is the particle speed and $L_c$ is the collision distance.

Because weakly collisional particles have large collision distances with
large diffusive Schwarz lengths $L_{SD}$ the non-baryonic dark matter
(possibly neutrinos) is the last material to fragment by self gravity and
not the first as assumed by CDM cosmologies.  The first structures occur as
proto-supercluster-voids in the baryonic plasma controlled by viscous and weak
turbulence forces, independent of diffusivity  ($[L_{SV}]_{plasma}
\le ct$, $D \approx \nu$).  The CDM
seeds postulated as the basis of CDMHCC never happened because $(L_{SD})_{NB}
\ge ct$ in the plasma epoch.

The baryonic matter is subject to large viscous
forces, especially in the hot primordial plasma and gas states 
existing when most
gravitational structures first formed.  The viscous forces per unit 
volume $\rho
\nu
\gamma L^2$ dominate gravitational forces $\rho^2 G L^4$ at small scales, where
$\nu$ is the kinematic viscosity and $\gamma$ is the rate of strain of the
fluid.  The forces match at the viscous Schwarz length
\begin{equation} L_{SV} \equiv (\nu \gamma /
\rho G)^{1/2},\end{equation}
  which is the smallest size for self gravitational condensation or
void formation in such a flow.  Turbulent forces may require even larger scales
of gravitational structures.  Turbulent forces $\rho \varepsilon^{2/3} L^{8/3}$
match gravitational forces at the turbulent Schwarz scale
\begin{equation}L_{ST} \equiv \varepsilon
^{1/2}/(\rho G)^{3/4},\end{equation}
  where $\varepsilon$ is the viscous dissipation rate of the turbulence. 
Because in the primordial plasma the viscosity and diffusivity are identical
and the rate-of-strain $\gamma$ is larger than the free-fall frequency $(\rho
G)^{1/2}$, the viscous and turbulent Schwarz scales
$L_{SV}$ and 
$L_{ST}$ will be larger than the diffusive Schwarz scale $L_{SD}$, from
(2), (3) and (4).  

Therefore, the criterion for structure formation in the plasma epoch is that
both $L_{SV}$ and $L_{ST}$ become less than the horizon scale $L_H = ct$. 
Reynolds numbers in the plasma epoch were near critical, with  $L_{SV} \approx
L_{ST}$.  From $L_{SV}< ct$ and (3), gravitational structures first formed when
$\nu < c^2 t (t^2 \rho G) \approx c^2 t$ at time $t \approx 10^{12}$ seconds
\citep{gib96}, well before $10^{13}$ seconds which is the time of plasma to gas
transition (300,000 years).  Because the expansion of the universe inhibited
condensation but enhanced void formation in the weakly turbulent plasma, the
first structures were proto-supercluster-voids. At $10^{12}$ s 
\begin{equation}
(L_{SD})_{NB} \gg L_{SV} \approx L_{ST} \approx 5 \times L_K \approx L_H = 3
\times 10^{20} \rm m, \end{equation} 
where $L_{SD}$ applies to the non-baryonic component and $L_{SV}$, $L_{ST}$,
and $L_{K}$ apply to the
baryonic component.

As proto-supercluster fragments formed, the voids filled with
non-baryonic matter by diffusion, inhibiting further structure formation by
decreasing the gravitational driving force.  The baryonic mass density
$\rho
\approx 2
\times  10^{-17}$ kg/$\rm
m^3$ and rate of strain
$  \gamma \approx 10^{-12}$ $\rm s^{-1}$ were preserved as hydrodynamic fossils
within the proto-supercluster fragments,  and also within
proto-cluster and proto-galaxy objects resulting from subsequent fragmentation
as the photon viscosity and
$L_{SV}$ decreased prior to the plasma-gas transition and photon decoupling
\citep{gib00}.  As shown in Eq. 5, the Kolmogorov scale $L_K \equiv [\nu^3
/\varepsilon ]^{1/4}$ and the viscous and turbulent Schwarz
scales at the time of first structure nearly matched the
horizon scale $L_H
\equiv ct \approx 3 \times 10^{20}$ m, freezing in the density, strain-rate, and spin magnitudes and
directions of the subsequent proto-cluster and proto-galaxy fragments of
proto-superclusters.  Remnants of the strain-rate and spin magnitudes and
directions of the weak turbulence at the time of first structure formation are
forms of fossil vorticity turbulence
\citep{gib99}. Thus, HGD explains galaxy spin
alignments and close angular associations with quasars 
without assuming Arp intrinsic red shifts and AGN quasar ejections.

The quiet condition of the primordial gas is  revealed by
measurements of temperature fluctuations of the cosmic microwave background 
radiation that show
an average $\delta T/T \approx 10^{-5}$ much too small for uninhibited strong 
turbulence to have
existed at that time of plasma-gas transition ($10^{13}$ s).  Turbulent plasma
motions were strongly damped by buoyancy forces at horizon scales after the first
gravitational fragmentation time 
$10^{12}$ s.  Viscous forces in the plasma are inadequate to explain the lack
of primordial turbulence ($\nu$
$ \ge 10^{30}$ m$^2$ s$^{-1}$ is required but, after $10^{12}$ s, $\nu \le 4
\times 10^{26}$, Gibson 2000). Thus the observed lack of strong turbulence
proves that large scale buoyancy forces and gravitational structure formation
must have begun in the plasma epoch.  The linear geometry and small $10^{20}$ m
scales of  proto-galaxy structures reflect the Kolmogorov scale and Nomura geometry
of weak plasma turbulence at transition to gas \citep{gib06,gs07,nom98}.  Observations
of CG and FG clusters \citep{men07} reveal proto-galaxy $L_N$ sizes and linear
geometries expected from turbulent plasma proto-galaxy-cluster fragmentation 
by the expansion of the universe.

\section{Stephan's Quintet:  HGD interpretation of an HST image}

Moles et al. 1997 summarize the data and dynamical status of SQ consistent with
standard CDMHC cosmology, proposing that the nearby NGC 7320C with $cz = 6.0
\times 10^{5}$ m/s (matching that of NGC 7318B) has possibly collided several
times with SQ members stripping their gas and central stars to form luminous
wakes and to  preserve their dynamical equilibrium, thus accounting for the
fact that 43 of the 100 members of the Hickson 1982 catalog of compact groups
contain discordant redshift members.  However, Gallagher et al. 2001 show from
their Hubble Space Telescope (HST) measurements that globular star clusters in
SQ are not concentrated in the inner regions of the galaxies as observed in
numerous merger remnants, but are spread over the SQ debris and surrounding
area.  We see no evidence of collisions or mergers in the HST images of SQ and
suggest the luminous wakes are not gas stripped from galaxy cores by collisions
but are new stars triggered into formation in the baryonic-dark-matter halo of
the SQ cluster as member galaxies are gently stretched away by the expansion of
space.  

According to HGD, galaxy mergers and collisions do not strip gas but
produce gas by evaporating the frozen hydrogen and helium of the planetary mass
objects which dominate the baryonic mass of galaxies.  The baryonic dark matter
is comprised of proto-globular-star-cluster (PGC) clumps of planetary-mass
primordial-fog-particles (PFPs) from hydro-gravitational-dynamics theory 
\citep{gib96} and quasar microlensing observations \citep{sch96}. Therefore from HGD the
cores of SQ galaxies are deficient in gas and YGCs because they have not
had mergers or collisions.  

Following standard CDMHC cosmology and N-body computer models,
galaxies and clusters of galaxies are formed by hierarchical collisionless
clustering due to gravity starting with
sub-galaxy mass CDM seeds condensed in the plasma epoch after the big bang.  The
Jeans 1902 gravitational condensation criterion rules out structures forming in
ordinary baryonic matter.  CDM seeds are diffusionally and tidally unstable from
hydro-gravitational-dynamics theory and their clustering to form galaxies is contrary to
observations
\citep{sa02}.  From HGD, both CDMHC
cosmology and the Jeans 1902 criterion are fundamentally incorrect and
misleading
\citep{gib00, gib06}.  CDM seeds cannot form because the CDM material is so weakly collisional.  CDM seeds cannot merge because tidal forces would cause rapid fragmentation of the merging seeds to scales of the fundamental particles.  

The unknown non-baryonic CDM material is enormously diffusive
compared to the H and He ions of the primordial plasma and cannot condense or
fragment gravitationally.  However, we can be sure structure formation
occurred in the plasma epoch because buoyancy within self
gravitational structure is the only  mechanism available to inhibit
turbulence.   Viscous forces were inadequate. 
Fully developed turbulence would have produced $\delta T/T
\approx 0.1$ values much larger than the $\delta T/T \approx 0.00001$ values
observed in numerous cosmic microwave background studies.  From HGD, structure formation first
occurred by gravitational fragmentation due to the expansion of space when
viscous and weak turbulence forces of the primordial plasma matched
gravitational forces at scales smaller than the horizon scale
$ct$, where $c$ is the speed of light and $t$ is the time after the big bang. 
The growth of structure was arrested by non-baryonic matter filling the voids
between baryonic fragments.  This HGD-cosmology and its application to
the interpretation of SQ is illustrated schematically in Figure 1ab.

In Fig. 1a at top left we see a fragmenting proto-supercluster ($10^{46}$ kg) of
the primordial plasma as it separates from other such fragments due to the
rapid expansion of the universe at the time of first gravitational structure
formation about 30,000 years ($10^{12}$ s) after the big bang
\citep{gib96}.  The scale is near the horizon scale $ct$ at that time $3 \times
10^{20}$ m with baryonic density $2 \times 10^{-17}$ kg/$\rm m^3$ and
non-baryonic density $
\approx 10^{-15}$ kg/$\rm m^3$  decreasing with time
and the non-baryonic matter (probably neutrinos) diffuses to fill the voids and
reduce the gravitational forces \citep{gib00}.  In Fig. 1a center proto-cluster
fragments form and separate, and on the right
proto-galaxies fragment just before the cooling plasma turns to gas at 300,000
years ($10^{13}$ s).  

The proto-galaxies preserve the density and spin of the
proto-supercluster as fossils of the primordial plasma turbulence
\citep{gib99}.  Their initial size is therefore about $3 \times 10^{19}$ m as
the plasma fragments with the inertial-vortex viscous scales and geometry
of weak turbulence. 
The gas proto-galaxies fragment into Jeans-mass  ($10^{36}$ kg) proto-globular-cluster (PGC)
dense clouds of ($10^{24}$ kg) primordial-fog-particles (PFPs) that cool,
freeze, and diffuse away from Nomura scale galaxy cores to form baryonic-dark-matter
(BDM) halos around galaxies and galaxy-clusters such as SQ.  The Jeans-mass is
relevant, but not for the reasons given by Jeans (1902).  Some galaxy-clusters
can be very slow in their separation due to crowding and frictional forces of
their BDM halos, as shown by the central galaxy cluster at the right of Fig.
1a. The BDM halo may reveal the history of galaxy mergers and separations
because strong tidal forces and radiation by galaxy cores trigger the formation
of stars and YGCs as they and their  halos move through each other's BDM halos,
leaving star wakes and dust wakes.

Fig. 1b shows schematically our interpretation of SQ based on HGD.  The five
galaxies are separated by distances inferred from Hubble's law and their red
shifts times the horizon distance $10^{26}$ m due to the stretching of space
along a thin tube of diameter $\approx 2
\times 10^{21}$ m oriented along the line of sight to the Trio.  The distance to
the line-of-sight tube entrance from earth is thus
$\approx 2.7
\times 10^{23}$ m for NGC 7320, with the exit and Trio at  $\approx 2.2
\times 10^{24}$ m.  NGC 7320 appears larger than the Trio members because it is
closer, consistent with the fact that it contains numerous
young-globular-clusters (YGCs) obvious in the HST images, but YGCs in the Trio are
barely resolved \citep{gal01}.  The tube in Fig. 1b is not to scale: the true
aspect ratio is that of a sheet of paper or a very long stick of uncooked
spaghetti.  By perspective,  about 1\% of the front face of the tube covers the
back face.

Figure 2 shows an HST image of Stephan's Quintet.  The trail of luminous
material extending southeast of NGC 7319 is interpreted from HGD as a star wake
formed as one of the galaxy-fragments of the original cluster moves away
through the baryonic-dark-matter (BDM) halo, triggering star formation until it
exits at the halo boundary marked by a dashed line.  Other star wakes in Fig. 2
are also marked by arrows.  These star wakes are similar in origin to the
filamentary galaxy VV29B of the Tadpole merger
\citep{gs03} and the ``tidal tails'' of the Mice and Antennae merging galaxies,
except that in SQ all the galaxies are seen to separate through each
other's halos rather than merge, contrary to the standard SQ \citep{mol97}
model.  

Two dust trails
are shown by arrows in the upper right of Fig. 2 that we interpret as star
wakes of the separation of NGC 7318B from NGC 7318A.  A similar dust trail is
interpreted from its direction as a star wake of NGC 7331 produced in the NGC
7319 BDM halo as it moved out of the cluster.  The luminous trail pointing
toward NGC 7320C is confirmed by  gas patterns \citep{gut02} observed from
broadband R measurements that suggest NGC 7320 has the same origin near NGC
7319.  An unidentified galaxy separated in the northern star
forming region, leaving over a hundred YGCs \citep{gal01} before  exiting the
BDM halo boundary (shown by the dashed line in the upper left of Fig. 2).

Details of the Hubble Space Telescope images of Stephan's Quintet (including
Fig. 2) can be found at the website for the July 19, 2001 STScI-2001-22 press
release\\ (http://hubblesite.org/newscenter/archive/2001/22/image/a).  The images
are described as ``Star Clusters Born in the Wreckage of Cosmic Collisions''
reflecting the large number of YGCs detected \citep{gal01} and the standard SQ
model \citep{mol97}.  

According to our HGD interpretation, none of the YGCs are
due to galaxy collisions or mergers.  All are formed in the BDM halos as the
galaxies gently separate with small transverse velocity along lines of
sight.  There were no cosmic collisions and there is no wreckage.  Numerous
very well resolved YGCs can be seen in the NGC 7320 high resolution image with
separations indicating numbers in the range
$10^{5} \-- 10^{6} $.  This suggests a significant fraction of the dark baryonic
matter in the halo of NGC 7320 has been triggered to form YGCs and stars as the
galaxy separated through both the dense BDM halo of the SQ Trio and the BDM
halo of its companion galaxy NGC 7331, also at $z=0.0027$ ( $2.7 \times 10^{23}$ m) 
separated northeast $3 \times 10^{21}$ m.   No such
concentration of YGCs can be seen in the SQ Trio galaxies, consistent with our
HGD interpretation that their distance is $\approx$8 times that of NGC 7320 as shown
in Fig. 1b.

\section{Stephan's Quintet:  HGD interpretation of R and $H_\alpha$ maps}

The present status of observations of Stephan's Quintet is well summarized 
by Gutierrez et al. 2002, including their deep broadband R and narrowband
$H_\alpha$ maps shown in Figure 3.  The R band map (their Fig. 1) with
sensitivity 26 mag arcsec$^{-2}$ extends to a wide range that includes NGC
7320C with the other SQ member galaxies.  A clear $H_\alpha$ bridge is shown
with red shift
$z=0.022$ corresponding to that of the SQ Trio to a sharp interface with
$z=0.0027$ material in NGC 7320, consistent with our interpretation that the
bridge was formed in the BDM halo of the SQ Trio by NGC 7320 as it emerged and
separated by the expansion of the universe along the line of sight, as shown by
the dashed arrow in Fig. 2.  

The solid arrow shown in Fig. 3 toward NGC 7320C suggests its emergence from
the SQ Trio BDM halo leaving the star wake shown by a corresponding arrow in
Fig. 2.  The mechanism of star wake production is that the frozen PFPs are in
meta-stable equilibrium within their PGCs.  Radiation from a passing galaxy 
causes evaporation of gas and tidal forces which together  increase the rate of
accretion of the PFPs to form larger planets and finally stars.  The size of
the stars and their lifetimes depends on the turbulence levels produced in the
gas according to HGD, Eq. (4).  Large turbulence levels produce large, short lived
stars.  The dust lane between NGC 7318A and its twin NGC 7318B suggests large
turbulence levels produced large stars and dust through
supernovas. A similar dust lane from NGC 7219 is in the general direction of
NGC 7331 and its companions, as indicated by the arrow in Fig. 2.

\section{Conclusions}

We conclude that Stephan's Quintet compact galaxy cluster (HCG92) 
with its highly anomalous redshifts is better described by
hydro-gravitational-dynamics (HGD) theory and cosmology \citep{gib96} 
than by the CDMHCC or the Arp AGN-ejected-galaxy intrinsic-redshift scenarios. According to HGD-cosmology, all the SQ galaxies formed 
in a linear cluster by gravitational
fragmentation of the primordial plasma just before photon decoupling and
transition to gas 300,000 years after the big bang.  None of the galaxies show
evidence of collisions or mergers. Such close alignments are
improbable by chance.  They remained stuck together for
12.9 billion years until 220 million years ago when the uniform expansion of
space in the universe finally overcame gravitation and the viscous
frictional forces of the cluster baryonic-dark-matter halo.  The BDM halo consists
of proto-globular-star-cluster (PGC) clumps of frozen primordial planets (PFPs) with
a large kinematic viscosity
\citep{gib06} because PGCs become weakly collisional as their planets freeze.

The nature of the baryonic-dark-matter halo is explained by HGD and supported by
the SQ observations.  At the plasma-gas transition the small, dense, proto-galaxy 
plasma-clouds turned to gas. From HGD \citep{gib96} the gas  fragmented at the
Jeans scale to form PGC ($10^{36}$ kg) clumps of ($10^{24}$ kg)
primordial-fog-particles (PFP planets), as shown in Fig. 1b, consistent with the conclusion
\citep{sch96} from quasar microlensing observations that the lens galaxy mass
is dominated by ``rogue planets likely to be the missing mass''.  

Some of the
PFPs near the proto-galaxy centers accreted to form stars and the luminous
galaxy cores. Most PFPs condensed and froze as the universe expanded and cooled
so their PGCs remained dark and gradually diffused away from the galaxy cores
to form BDM galaxy halos, and some diffused further to form cluster
baryonic-dark-matter (BDM) halos.   The Stephan Quintet cluster BDM halo
boundaries are revealed by the separation of the SQ galaxies as star wakes, as
shown in Fig. 2.  The SQ BDM halo radius is only
$\approx 2
\times 10^{21}$ m,  compared with the BDM halo radius of the
Tadpole galaxy $\approx 5 \times 10^{21}$ m as shown
by HST/ACS images with the star wake of the merging galaxy \citep{gs03}.

Our HGD interpretation of SQ solves the long standing mystery of its anomalous
red shifts \citep{bur61}.  Rather than an explosive expansion or intrinsic red
shifts of the SQ galaxies ejected by the same parent \citep{arp73} we suggest 
that a uniform expansion of the universe stretched the SQ
galaxies along a line of sight because of perspective and
small transverse velocities resulting from BDM halo gas friction and their
sticky beginnings,  as shown in Fig. 1b.  The common point of origin of the SQ
galaxies is confirmed by gas trails in recent R and
$H_\alpha$ maps, as shown in Fig. 3
\citep{gut02}. Highly discordant red shifts often observed for
aligned quasars and AGN galaxies 
\citep{hoy00} are thus explained using the conventional physics of HGD.

From the present study, the interpretation of compact groups [CG] and fossil groups [FG] of galaxies as mergers  \citep{men07} should be reversed.  These are not merging but separating galaxy clusters.

\acknowledgments

\clearpage

\begin{figure}
        \epsscale{0.5}
        \plotone{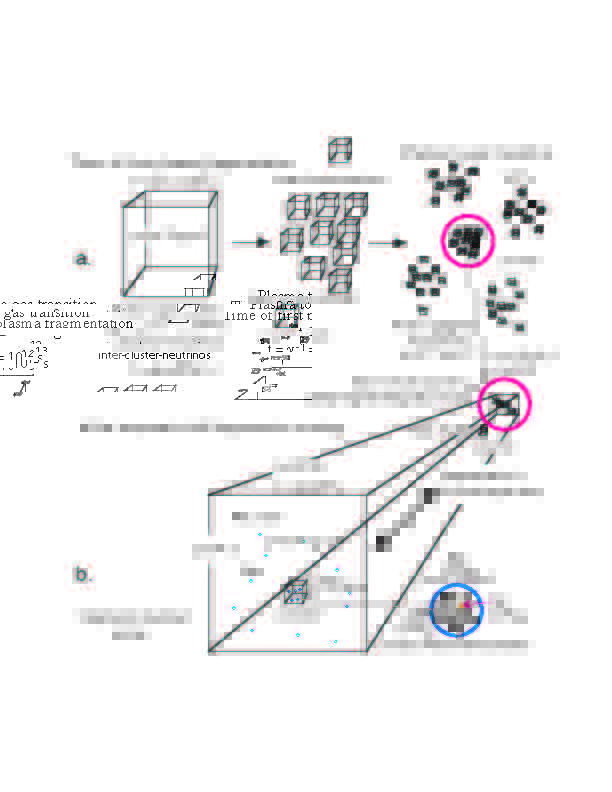}
        \caption{a.  According to hydro-gravitational-dynamics
\citep{gib96}, proto-superclusters (left) fragment to proto-clusters (center)
which fragment to form proto-galaxies during the super-viscous plasma epoch
\citep{gib00}. 
Compact galaxy clusters such as Stephan's Quintet occur in this cosmology
when dispersal of the cluster by the expansion of the universe is delayed by
frictional forces; eg., the central cluster of galaxies on the right.  b. 
Galaxies of the fragmented SQ cluster remain along a line of sight to the SQ
Trio because of their small transverse velocities, reflecting their sticky
beginnings.}
       \end{figure}

\begin{figure}
        \epsscale{0.5}
       \plotone{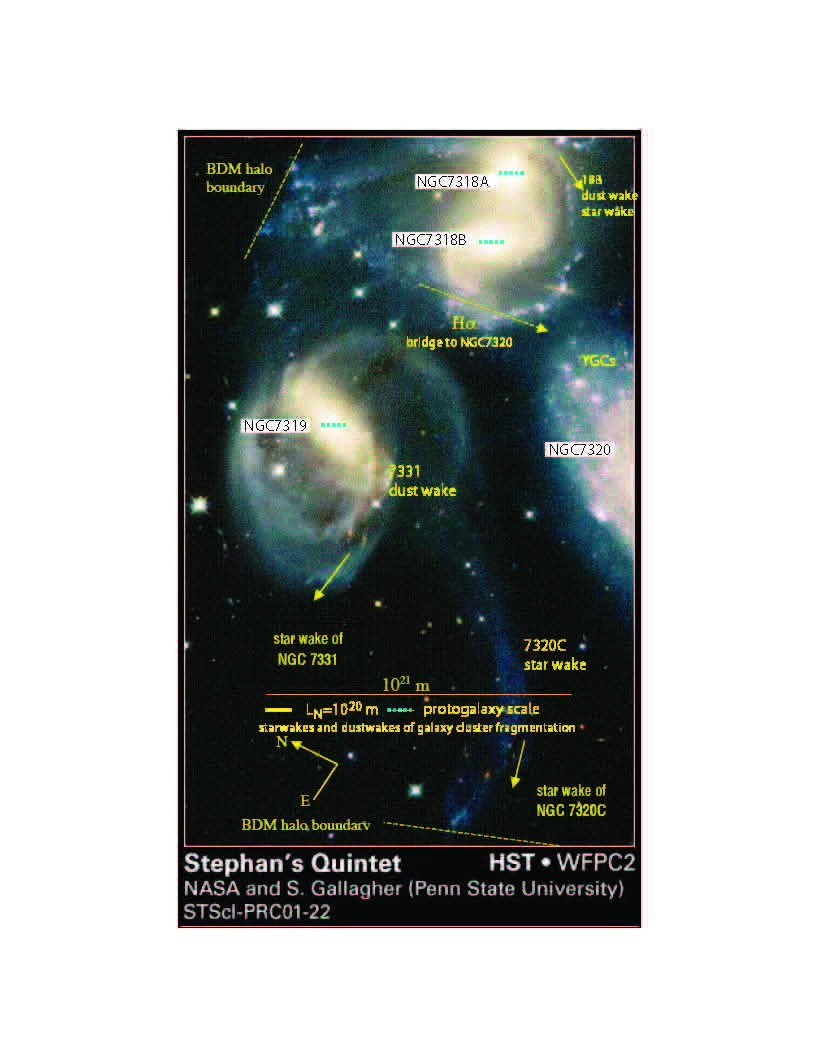}
        \caption{Hubble Space Telescope image of Stephan's Quintet. Dust and
star wakes (arrows) are produced as SQ related galaxies gently separate from
each other through the Trio cluster baryonic-dark-matter (BDM) halo of PGCs and PFPs,
triggering star formation.  Star wakes of mergers and collisions are not
observed.  The $H_\alpha$ gas bridge proves NGC7320 is not a chance intruder.  BDM haloes
form by diffusion of PGC clumps of dark matter planets from Nomura scale $L_N$ galaxy cores that reflect the size of plasma protogalaxies (Fig. 1a).  YGCs can be resolved in NGC7320
because it is closer than the Trio, contrary to the Arp hypothesis that SQ galaxies 
are near and ejected by  parent NGC7331 with intrinsic redshifts \citep{arp73, arp98}.}
       \end{figure}

\begin{figure}
        \epsscale{0.5}
        \plotone{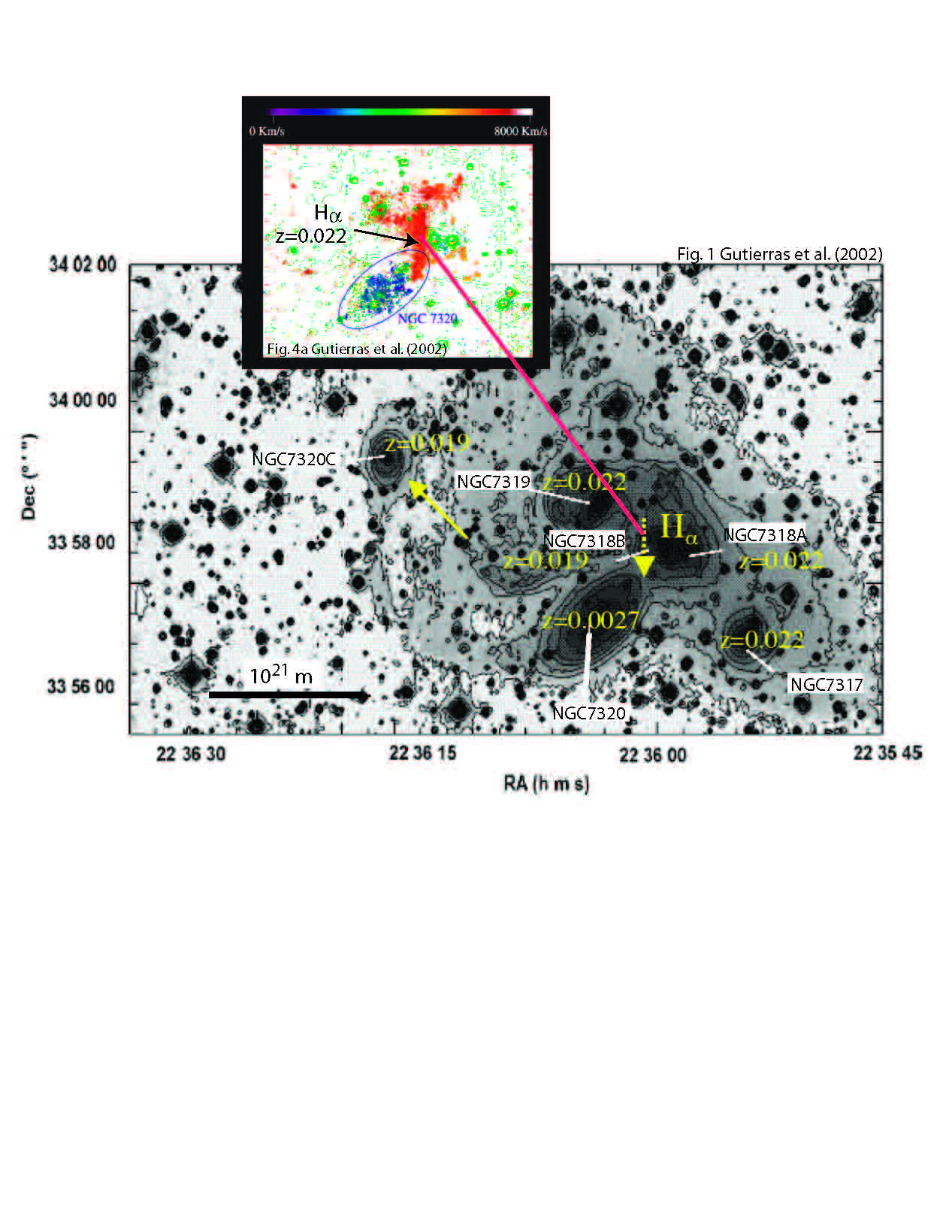}
        \caption{Contour R map of SQ \cite{gut02} showing connections between
SQ galaxies and NGC 7320C to the East (left), and NGC 7320 to the South
(bottom).  The $H_\alpha$ bridge is at the red shift 0.022 of the SQ Trio, and
shows a sharp transition to $z=0.0027$ for NGC 7320 \citep{gut02} consistent
with HGD cosmology where fragmented SQ galaxies are gently stretched into a thin
pencil by the expansion of the universe,  Fig. 1b. }
       \end{figure}


\begin{thebibliography}{}

\bibitem[Aharonian et al. 2006]{ah06} Aharonian et al. 2006, Nature, 440, 1018

\bibitem[Arp 1973]{arp73} Arp, H. 1973, \apj, 183, 411

\bibitem[Arp 1998]{arp98} Arp, H. 1998, \apj, 496, 661

\bibitem[Burbidge and Burbidge 1961]{bur61} Burbidge, E. M., \& Burbidge, G. R.
1961, ApJ, 134, 244

\bibitem[Burbidge 2003]{bur03} Burbidge, G. R.
2003, ApJ, 586, L119

\bibitem[Bershadskii and Sreenivasan 2002]{bs02} Bershadskii, A., and K.R.
Sreenivasan, 2002, Phys. Lett. A, 299,  149

\bibitem[Gallagher et al. 2001]{gal01} Gallagher, S. C., Charlton, J. C.,
Hunsberger, S. D., Zaritsky, D., \& Whitmore, B. C. 2001, AJ, 122, 163

\bibitem[Gibson 1996]{gib96} Gibson, C. H.
1996, Appl. Mech. Rev., 49, 299, astro-ph/9904260

\bibitem[Gibson 1999]{gib99}Gibson, C. H. 1999, J. of Mar. Systems,
21, 147, astro-ph/9904237
 
\bibitem[Gibson 2000]{gib00} Gibson, C. H.
2000, J. Fluids Eng., 122, 830,
astro-ph/0003352.

\bibitem[Gibson 2001]{gib01} Gibson, C. H. 2001, Proc. ICME 2001, Vol. 1,
BUET, 1, astro-ph/0110012

\bibitem[Gibson 2006]{gib06} Gibson, C. H. 2006, J. Appl. Fluid Mech. 
(in press 2008, 2(1), 1-8, www.jafmonline.net), astro-ph/0606073v3

\bibitem[Gibson 2004]{gib04} Gibson, C. H. 2004, Flow, Turbulence and Combustion, 72, 161Ð179

\bibitem[Gibson 2005]{gib05} Gibson, C. H. 2005, Combust. Sci. and Tech., 177,  1049-1071

\bibitem[Gibson 2006]{gib06} Gibson, C. H. 2006, astro-ph/0606073

\bibitem[Gibson \& Schild 2003]{gs03} Gibson, C. H. \& Schild, R. E. 2003, astro-ph/0210583v2

\bibitem[Gibson \& Schild 2007]{gs07} Gibson, C. H. \& Schild, R. E. 2007, astro-ph/0701474

\bibitem[Gutierrez et al. 2002]{gut02} Gutierrez, C. M., Lopez-Corredoira, M.,
Prada, F. \& Eliche, M. C. 2002, \apj, 579, 592

\bibitem[Hickson 1982]{hic82} Hickson, P. 1982, \apj, 255, 382

\bibitem[Hoyle et al. 2000]{hoy00} Hoyle, F., Burbidge, G., \& Narlikar, J. V.
2000, A Different Approach to Cosmology, Cambridge U. Press

\bibitem[Jeans 1902]{jns02} Jeans, J. H. 1902,
Phil. Trans. R. Soc. Lond. A, 199, 1

\bibitem[Mendes de Oliveira  \& Carrasco 2007 ]{men07}Mendes de Olivieira, C. \& Carrasco, E. R. 2007, 
arXiv:0710.3347v1[astro-ph], accepted ApJL

\bibitem[Moles et al. 1997]{mol97} Moles, M., Sulentic, J. W., \& Marquez, I.
1997, \apj, 485, L69

\bibitem[Nomura \& Post 1998]{nom98} Nomura, K. K. \& Post, G. K. 1998, JFM, 377, 65-97

\bibitem[Schild 1996]{sch96} Schild, R. 1996,
\apj, 464, 125

\bibitem[Sand et al. 2002]{sa02} Sand, D. J., Treu, T, \&
Ellis, R. S. 2002, ApJ, 574, L129


\end{thebibliography}
\end{document}